\documentclass[9pt,twocolumn,twoside]{pnas-new}
\templatetype{pnasresearcharticle} % Choose template 
\title{Terminating spiral waves with a single designed stimulus: Teleportation as the mechanism for defibrillation}

\author[a,1]{Noah DeTal}
\author[a]{Abouzar Kaboudian} 
\author[a]{Flavio Fenton}

\affil[a]{School of Physics, Georgia Institute of Technology, Atlanta, GA 30332}

\leadauthor{DeTal} 

\significancestatement{Many chemical and biological systems can sustain complex spiral wave dynamics. In the heart, spiral waves of electrical activity induce deadly arrhythmias and must be eliminated with a large, system-wide perturbation to restore a healthy rhythm. However, the high-energy shocks required for defibrillation therapies are very painful and can even damage heart tissue. In this study, we demonstrate a new method for eliminating spiral waves with the minimal possible stimulus. To do this, we show how a localized perturbation can be designed to \textit{“teleport”} pairs of oppositely rotating spirals close together such that they mutually annihilate. This newly identified mechanism is applicable to any excitable system, but for the heart has the potential to lead to more efficient defibrillation strategies.}

\correspondingauthor{\textsuperscript{1} E-mail: nddetal@gmail.com}


\begin{abstract}
We demonstrate a universal mechanism for terminating spiral waves in excitable media using an established topological framework. This mechanism dictates whether high- or low-energy defibrillation shocks succeed or fail. Furthermore, this mechanism allows for the design of a single minimal stimulus capable of defibrillating, at any time, turbulent states driven by multiple spiral waves. We demonstrate this method in a variety of computational models of cardiac tissue ranging from simple to detailed human models. The theory described here shows how this mechanism underlies all successful defibrillation and can be used to further develop existing and future low-energy defibrillation strategies.
\end{abstract}

\dates{This manuscript was compiled on \today}
\doi{\url{www.pnas.org/cgi/doi/10.1073/pnas.XXXXXXXXXX}}

\begin{document}

\maketitle
\thispagestyle{firststyle}
\ifthenelse{\boolean{shortarticle}}{\ifthenelse{\boolean{singlecolumn}}{\abscontentformatted}{\abscontent}}{}

% If your first paragraph (i.e. with the \dropcap) contains a list environment (quote, quotation, theorem, definition, enumerate, itemize...), the line after the list may have some extra indentation. If this is the case, add \parshape=0 to the end of the list environment.
\dropcap{S}piral waves have been shown to exist in many chemical \cite{winfree1972spiral} and biological \cite{tyson1989spiral} systems, with many of these supporting spiral-wave-mediated turbulence \cite{tsuji2019spirals}. Of particular clinical significance are the spiral waves of electrical activity that underlie tachycardia and fibrillation in the heart \cite{ricknat, cherry2008visualization}. Even before spiral waves were recognized as drivers of tachycardia and fibrillation in the 1990s \cite{davidenko1992stationary}, termination of these deadly arrhythmias was shown possible by large electric shocks as early as 1899. However, it was not until 1947 that defibrillation was shown to be successful in the clinic \cite{eisenberg1998defibrillation}. Defibrillation usually requires electric shocks on the order of 5-7 Joules for internal and 200-300 Joules for external devices \cite{fenton2009termination}. These high-energy shocks are very painful and can damage heart tissue but are required to terminate all reentrant waves and prevent initiation of new spiral waves \cite{efimov1998virtual}.

While currently there are no clinically viable low-energy defibrillation strategies, several new methods have been developed and tested computationally and experimentally.  One consists of a multi-stage series of low-energy pulses intended to subsequently unpin and remove reentrant spiral waves \cite{li2011low}. Another example is the low-energy antifibrillation pacing (LEAP) method \cite{leap}, which uses a series of fast pulses at a frequency close to the dominant frequency of the arrhythmia to eliminate spiral waves \cite{fenton2009termination}. Both of these methods utilize the mechanism of \textit{virtual electrodes}---tissue activation driven by the heart’s natural heterogeneities \cite{efimov1998virtual,fenton2009termination}. Other studies have used concepts from nonlinear dynamics to determine optimal defibrillation stimulus timing and strength by calculating phase \cite{rickgray} and isostable \cite{danwilson} resetting curves. However, neither of these methods directly address the topological spatiotemporal coupling responsible for spiral waves and both require detailed spatial information of the system.  In this article, we demonstrate how to design a single minimal stimulus from topological considerations and phase space values to automatically eliminate reentrant spiral wave singularities in any complex fibrillating state.

\section*{Defining level set contours and phase singularities}
To characterize the complex spatiotemporal dynamics of an excitable system, regions in space are designated locally as excited, refractory, or recovered. For a multi-component system $\vec{u}(\vec{x},t)$, indicator functions $f(\vec{u})$ and $g(\vec{u})$ are defined such that regions where $f>0$ are excited and regions where $g>0$ are refractory \cite{topological}. The level sets $f=0$ and $g=0$ then define one-dimensional contours marking the boundaries of excited and refractory regions, respectively. Excited or refractory fronts can be differentiated from backs on a given zero-level set by the sign of the opposite indicator function \cite{topological, keener}. This results in a set of four distinct contours as defined in Table \ref{tab:conventions}. 

\begin{figure}[ht!]
    \centering
    \includegraphics[scale=0.43]{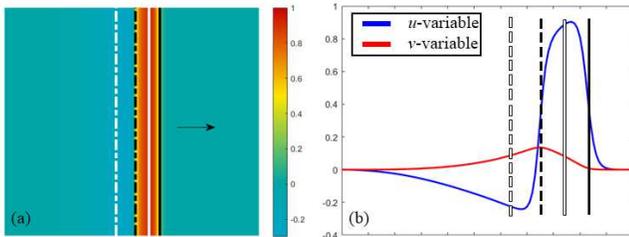}
    \vspace*{-5mm}
    \caption{(a) Rightward-traveling pulse in the two-variable FHN model on a 2D domain. Excited and refractory contours are superimposed following the conventions in Table \ref{tab:conventions}. (b) One-dimensional horizontal cross section of (a) showing both the $u$ (blue) and $v$ (red) variables. Level set contours are overlaid at positions from (a) indicating their corresponding values.} 
    \label{fig:wave}
\end{figure}

For two-variable reaction-diffusion systems such as the FitzHugh-Nagumo (FHN) \cite{fitzhugh1961impulses} and Karma \cite{karma1993spiral} models, $\vec{u}=\{u,v\}$ and obeys
\begin{equation}
    \begin{split}
        \frac{\partial u}{\partial t} &= h(u,v)+D\nabla^2 u \\
        \frac{\partial v}{\partial t} &= w(u,v).
    \end{split}
\end{equation}
In this case, the indicator functions $f$ and $g$ can simply be taken to be
\begin{equation}
    f = u-u_\textrm{th}, \quad \quad g = v-v_\textrm{th}
    \label{fg}
\end{equation}
for some threshold values $u_\textrm{th},v_\textrm{th}$. Other common choices of $f$ and $g$ include the system variables' time derivatives \cite{topological} or time-delayed values \cite{experimentalist}, local curvature \cite{curvature}, and normal velocity \cite{fenton1998vortex}. Each choice is topologically equivalent and gives comparable results \cite{topological, experimentalist}. Gurevich et al. have shown how to construct valid combinations using only voltage data in both simulations and optical mapping experiments of fibrillation with multiple waves even when there is noise in the data \cite{robust,robustexp}. In their implementations, $f$ and $g$ are determined by the extrema and inflection points of the local voltage and do not require measurement of the underlying gating currents. Figure \ref{fig:wave} shows the level set contours using (\ref{fg}) for a rightward-traveling pulse in the FHN model on a 2D domain (see SI Appendix, Supplementary Text for models and parameters). Such a pulse displays how local excitation must pass through excited front, refractory front, excited back, and refractory back in sequence before returning to rest.

\begin{table}
    \centering
    \caption{\label{tab:conventions} Definitions and plotting conventions for the four topological level set contours.}
    \begin{tabular}{ |l|l|l| } 
     \hline
     Excited front & $f=0,g<0$ & Denoted by solid black line \\ 
     \hline
     Refractory front & $g=0,f>0$ & Denoted by solid white line \\ 
     \hline
     Excited back & $f=0,g>0$ & Denoted by dashed black line \\ 
     \hline
     Refractory back & $g=0,f<0$ & Denoted by dashed white line \\ 
     \hline
    \end{tabular}
\end{table}

An important use of level set contours is the direct identification of spiral wave tips. These are isolated points at the center of spiral waves where wavefront and waveback intersect to form a continuous source of reentrant excitation. Spiral tips are most commonly described as phase singularities in a complex order parameter given by 

\begin{equation}
    Z(\vec{u}) = r \textrm{e}^{\textrm{i}\theta}=f(\vec{u})+g(\vec{u})\textrm{i}.
    \label{order}
\end{equation}

in analogy with homotopy theories used in condensed matter physics \cite{mermin}. The spatial line integral of the phase $\theta$ around a phase singularity (known as the topological charge) is equal to $\pm 1$ with the sign determining the corresponding spiral's chirality  \cite{experimentalist}. Equivalently, the amplitude $r$ vanishes exactly at a phase singularity, that is, when $f=g=0$ and the level set contours of Table \ref{tab:conventions} intersect. From Eq. \ref{fg}, this implies that $u=u_\textrm{th},v=v_\textrm{th}$ at the singularity. The chirality can be obtained from the level sets by calculating the sign of $\hat{z}\cdot \left( \nabla f \times \nabla g\right)$ at the singularity.

\begin{figure}[ht!]
    \centering
    \includegraphics[scale=0.32]{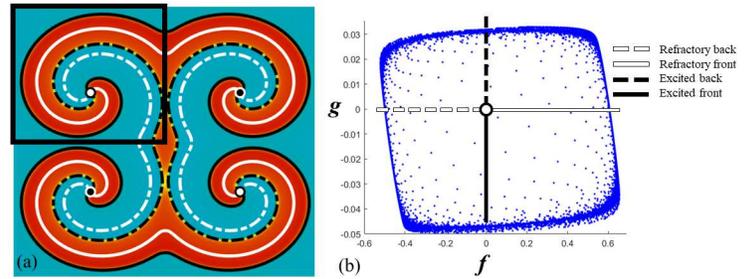}
    \caption{(a) Single spiral wave in the two-variable FHN model on a 2D domain with reflective boundary conditions  (dark frame). The extended domain shows how mirrored spiral waves allow for conservation of topological charge. Chirality is indicated with white (black) circles for counterclockwise (clockwise) spiral tips. (b) Phase space representation of the spiral wave shown in (a). Blue dots show the $f-g$ phase space position at every point in the physical 2D domain. Level sets are shown using conventions in Table \ref{tab:conventions}.}
    \label{fig:spiral}
\end{figure}

The framed region in Figure \ref{fig:spiral}(a) shows the contours and phase singularity for a single spiral wave in the FHN model. Figure \ref{fig:spiral}(b) shows the same spiral but in the phase space of the $f$ and $g$ indicator functions. The periodic behavior of spiral rotation manifests as a dense collection of points forming a closed loop. Because the origin is interior to this loop, the order parameter acquires a singularity in the presence of a spiral wave. 

The topological properties of phase singularities have long been understood to play a major role in the initiation and persistence of spiral-wave-mediated turbulence responsible for cardiac arrhythmias \cite{ricknat}. Winfree used topological arguments about the phase of the complex order parameter to show how singularities can be created or destroyed by particular stimuli \cite{winfree, winfreetop}. A significant result is that (excluding boundary interactions), singularities can only be created or destroyed in pairs of opposite chirality. This principle is commonly referred to as conservation of topological charge and follows simply from the continuity of $\vec{u}$ and the indicator functions $f,g$ \cite{topological,winfreetop,davidsen2004topological,pertsov2000topological}. Additionally, every singularity is uniquely connected to a singularity of opposite chirality by each of the four contours. However when multiple spirals exist, pairing between singularities will change over time as the spirals rotate.

Topological charge is not conserved on domains with zero-flux (or reflective) boundary conditions, which are used in cardiac simulations for conservation of charge \cite{noflux}. However, since the domain is effectively mirrored by such boundary conditions, as shown in Figure \ref{fig:spiral}(a), the extended domain does conserve topological charge. In this construction spirals with contours terminating at boundaries are connected to a mirrored spiral with opposite chirality. It follows that in order to eliminate spiral wave turbulence (or fibrillation), it is necessary to annihilate every singularity by merging it with a singularity of opposite chirality inside the domain or with its mirror image at a domain boundary. In the next section we show how this can be accomplished with an excitatory stimulus using only the topological structure of the level set contours.

\section*{Moving and terminating singularities with a single designed stimulus}
While phase and level set descriptions are topologically equivalent and are explicitly related through Equation (\ref{order}), each is suited to different systems. In oscillatory media, phase has a natural definition in terms of the underlying limit cycle and is thus the natural object to study \cite{kuramoto}. The excitable-refractory paradigm of excitable media, however, is better represented by the level set contours. In particular, they can be used to understand how stimulus leads to the creation or annihilation of spiral wave singularities.

\begin{figure}[ht!]
    \centering
    \includegraphics[scale=0.22]{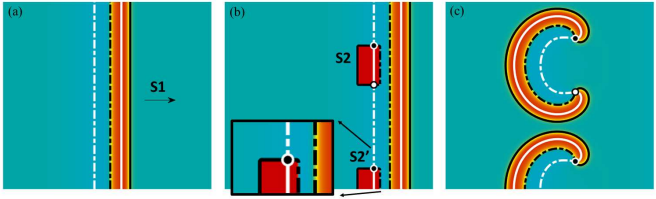}
    \caption{Creating spiral waves with the S1-S2 protocol. (a) A rightward-traveling pulse generated by exciting the left boundary with stimulus S1. (b) An opposite pair and an isolated singularity are created by applying localized S2 and S2' stimuli to the refractory back of the S1 pulse. As shown in the inset, the singularities are generated at the convergence of all four level set contours. (c) The singularities persist and form stable spiral waves that act as continuous sources of excitation. A complete video of this process is available in SI Appendix, Movie S1.}
    \label{fig:s1s2}
\end{figure}

A classic procedure for initiating a pair of spiral wave singularities via excitatory stimulus is the S1-S2 protocol \cite{s1s2,s1s22}, a variation of Winfree's pinwheel experiment. First, a stimulus S1 is applied to generate a traveling pulse. A second localized stimulus S2 is then applied behind the pulse. The excited wavefront of S2 intersects with the refractory back of the S1 pulse and generates a pair of singularities that develop into spiral waves as illustrated in Figure \ref{fig:s1s2}. Because the simulation uses zero flux boundary conditions, a stimulus S2' localized near the boundary may produce a single spiral singularity and violate conservation of topological charge.

Winfree originally explained the S1-S2 generation of singularities in terms of the continuous gradient of phase produced in the complex order parameter \cite{winfree}. For excitable media, however, the contours tell a simpler story. By definition, the refractory back contour separates regions which will respond to stimulus by exciting from regions which will be unexcited. This provides a mechanism for creating unidirectional propagation, as excitation can only travel away from the refractory region---left in the case of Figure \ref{fig:s1s2}(b). From Figure \ref{fig:wave}, we know that the excited portion of a traveling pulse contains an excited front, refractory front, and excited back. Stimulating a refractory back contour thus replaces it with this sequence of three contours. However, since the stimulus is localized, only a finite segment of the refractory back contour is replaced. At the boundary of the stimulus, the refractory back contour must connect with this sequence and therefore produce a singularity. This is illustrated in the inset of Figure \ref{fig:s1s2}(b).

\begin{figure*}[ht!]
    \centering
    \includegraphics[scale=0.35]{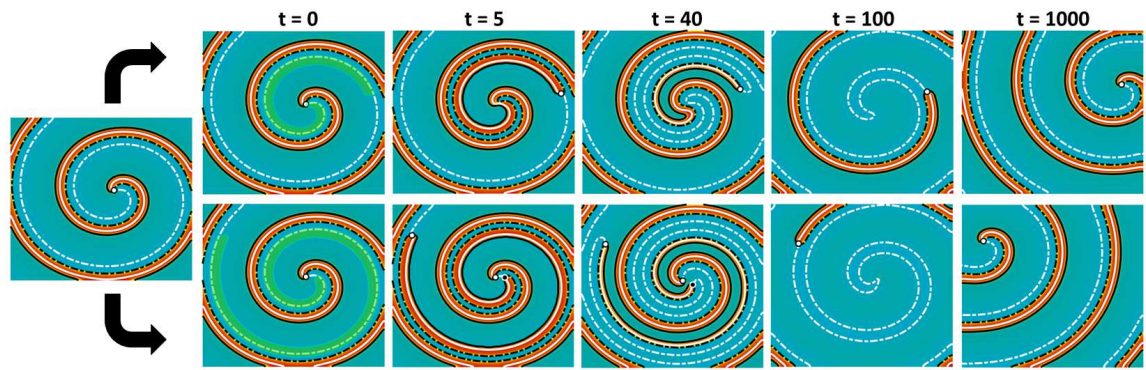}
    \caption{Two examples of teleporting a spiral wave to a new location. The top row shows application of a uniform stimulus along the singularity's refractory back and subsequent time lapse. The spiral wave tip instantaneously appears in the new position near the right boundary and begins to rotate. After a few rotations, the spiral arms are regenerated and fill the domain. The bottom row shows a discontinuous stimulus that starts close to the original spiral and terminates at a distant position close to the left boundary. In this case, two new spiral waves with opposite chirality are created. The first appears close to the original but has opposite chirality and thus annihilates with it. This leaves the second, teleported spiral wave with the same chirality as the original. As in the other case, after a few rotations all arms re-appear. Time displayed corresponds to dimensionless integration time. Complete videos of both examples are available in SI Appendix, Movies S2 and S3.}
    \label{fig:teleport}
\end{figure*}

The contour framework can also be used to design stimuli capable of instantaneously moving spiral wave tips arbitrary distances across the domain. When the refractory back of an existing singularity is uniformly stimulated, the singularity reforms at the end of the stimulus where once again the refractory back intersects with the newly formed excited front. The result is that the singularity is instantaneously moved, or \textit{``teleported''} along the refractory contour. This is demonstrated in Figure \ref{fig:teleport}, where a single spiral is teleported to two different locations, one near and one far, corresponding to different designed stimuli. During the course of a spiral's rotation, the refractory back will pass through every point in the domain, and so spirals can be teleported to any desired location by waiting an appropriate amount of time before stimulating. The first example in Figure \ref{fig:teleport} shows how a uniform designed stimulus starting near the initial spiral wave tip moves the spiral wave instantaneously to the opposite end of the stimulus. In the second example, the stimulus is 
not continuous and also terminates much further away. At the break in the stimulus, two new spiral waves of opposite chirality are created---one close to the original spiral wave tip, and one at the far end of the stimulus. The original singularity quickly annihilates with the newly created one of opposite chirality, leaving only the distant spiral wave. This latter example is similar to the destructive teleportation of science fiction and 3D remote printing \cite{mueller2015scotty}, which function by destroying the original object and creating an exact copy in a new location. 

\begin{figure}[ht!] 
    \centering
    \includegraphics[scale=0.22]{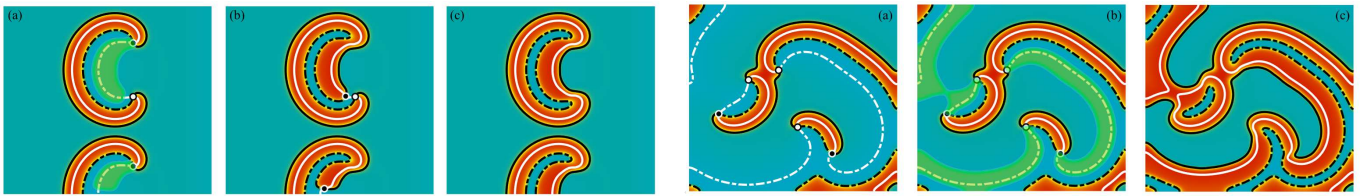}
    \caption{Terminating spiral waves with a single stimulus. (a) Regions covering the refractory back contours are selected for stimulus (light green). (b) Applying the stimulus ``teleports" the existing clockwise singularities along the refractory contours toward their counterparts. (c) The opposite pairs of singularities mutually annihilate and the contours reform without intersection. A complete video of this process is available in SI Appendix, Movie S1.}
    \label{fig:s3}
\end{figure}

While the S1-S2 protocol illustrates clearly how singularities can be created, it is not immediately obvious how to reverse this procedure in order to eliminate them. However, teleportation provides an important clue. Consider the new stimuli shown in Figure \ref{fig:s3}(a) superimposed on the configuration from Figure \ref{fig:s1s2}(c). These stimuli cover the clockwise singularities and a significant portion of their refractory back contours. The result of the stimuli, shown in Figure \ref{fig:s3}(b), is that the singularities are instantaneously teleported along the contour to the end of the stimulus. This brings the paired spiral very close to its partner and the isolated spiral very close to its mirrored image at the boundary. The nearby pairs then mutually annihilate due to their proximity. If the stimulus is allowed to cover the entire refractory back, the existing singularities will be teleported together and eliminated instantaneously as in Figure \ref{fig:s3}(c). This whole process can be understood from the contour topology in the same way as the S1-S2 protocol. Stimulating the refractory back once again replaces it with the excited front, refractory front, excited back contour sequence. This time, however, it causes the contours to disconnect (Figure \ref{fig:s3}(c)), thereby removing the singularities at either end of the contour.

Because every singularity is connected to a singularity of opposite chirality, a single stimulus targeting the refractory back can always be designed to teleport and eliminate all internally paired and boundary paired singularities. Moreover, excitation of the full refractory back contour can be taken as the necessary and sufficient condition for total defibrillation; the only way to remove singularities is to modify their topology in this manner. It should also be noted that the same topological arguments allow defibrillation using instead \textit{de}exciting stimuli to hyperpolarize the refractory \textit{front} \cite{keener}.

Previous theoretical analyses have reported a defibrillation threshold for domain-wide stimulation \cite{ideker, pumirkrinsky}. In that case, even refractory regions of tissue must be excited such that propagating fronts only contract and dissipate \cite{maxwell}. In our proposed method, only a small fraction of the domain needs stimulation. Additionally, since only the recovered region (which is readily excitable) of the refractory back needs to be excited, the stimulation strength can be well below the defibrillation threshold. 

In practice, the minimal defibrillating stimulus is constructed by applying a current $I_\textrm{stim}$ to the voltage variable $u$ wherever $f<0$ and $|g|<g_\textrm{th}$. The zero threshold $g_\textrm{th}$ gives the stimulus a finite thickness about the refractory back contour. If $I_\textrm{stim}$ or $g_\textrm{th}$ are too small, the stimulus will fail to permanently remove every singularity; although they may disappear initially, pairs of singularities can spontaneously reform and persist.

\begin{figure}[ht!] 
    \centering
    \includegraphics[scale=0.22]{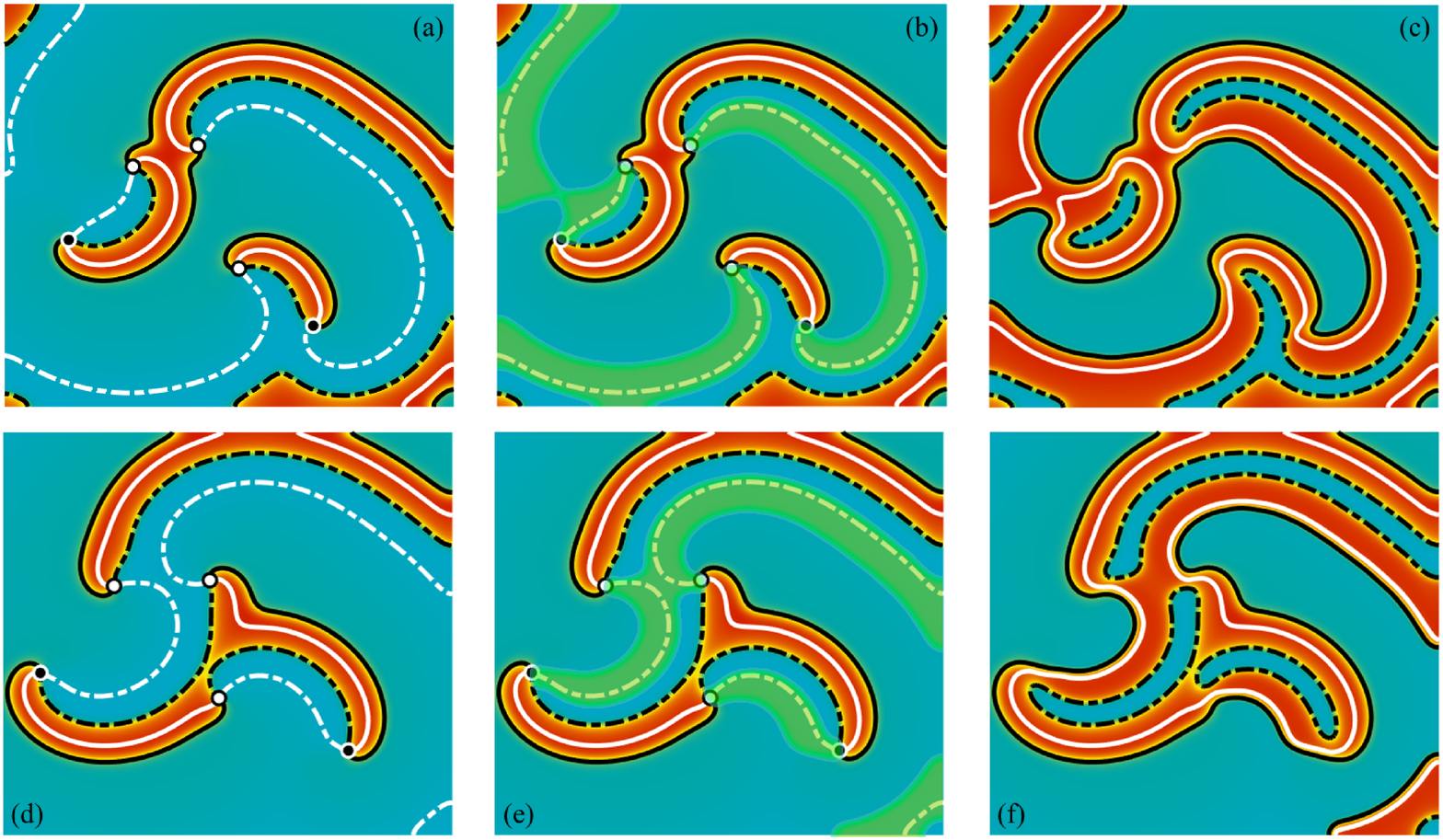}
    \caption{(a) A complex multi-spiral state in the FHN model. (b) Stimulus pattern designed to teleport all pairs of singularities along their refractory contours. (c) Successful defibrillation immediately after stimulus. (d) The complex state from (a) evolved in time such that the two lower spirals are connected. (e) Altered stimulus pattern for the configuration in (d). (f) Successful defibrillation of (d). A complete video of this process is available in SI Appendix, Movie S4.}
    \label{fig:FHNmulti}
\end{figure}

\begin{figure}[!ht]
    \centering
    \includegraphics[scale=.22]{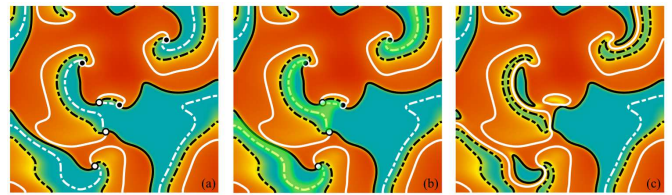}
    \caption{(a) Fibrillating spiral breakup in the Karma model. (b) Stimulus pattern designed to teleport all pairs of singularities along their refractory contours. (c) Successful defibrillation immediately after stimulus. A complete video of this process is available in SI Appendix, Movie S5.}
    \label{fig:KarmaS}
\end{figure}

Figure \ref{fig:FHNmulti} and Figure \ref{fig:KarmaS} show how the designed stimuli successfully teleport and remove all singularities in complex, multi-spiral states regardless of whether singularities are internally paired or connected to the domain boundary. In the Karma model, success is obtained despite spiral waves in the fibrillating state continuing to dynamically form and terminate \cite{karma1993spiral}.  Some configurations contain refractory backs not connected to singularities (e.g. upper left corner in Figure \ref{fig:FHNmulti}(a)). Stimulation of these contours is not required for defibrillation and is omitted in Figure \ref{fig:KarmaS}, further decreasing the fraction of the domain required to be stimulated. As mentioned previously, connections between pairs of singularities can also switch over time. Figure \ref{fig:FHNmulti}(d) shows an alternate configuration obtained a quarter of rotation after Figure \ref{fig:FHNmulti}(a) in which the two lower spirals have become connected. Defibrillation is successful in both models (FHN and Karma) despite the stimuli covering less than 10\% of the domain in the latter case. 

\begin{figure}[!ht]
    \centering
    \includegraphics[scale=.22]{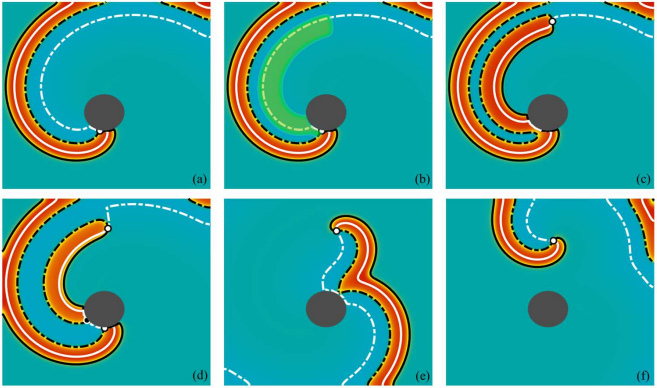}
    \caption{(a) A single spiral wave pinned to an obstacle in the FHN model. (b) Stimulus pattern designed to teleport the spiral away from the obstacle. (c) Immediate result of stimulation. The spiral singularity is instantaneously teleported to the edge of the stimulus. (d) Excitation at the obstacles creates a new pair of singularities. (e) The new singularity pair meets and annihilates on the opposite side of the obstacle. (f) The original spiral wave is completely unpinned. A complete video of this process is available in SI Appendix, Movie S6.}
    \label{fig:unpin}
\end{figure}

Teleportation can also unpin spiral waves bound to an obstacle. Figure \ref{fig:unpin} shows this process for a single spiral in the FHN model. A stimulus along the refractory back instantaneously teleports the singularity off the obstacle (Figure \ref{fig:unpin}(b) and (c)). However, the stimulus along the obstacle generates a new pair of singularities a short time later (Figure \ref{fig:unpin}(d)). These singularities rotate around the obstacle until they meet and annihilate on the opposite side (Figure \ref{fig:unpin}(d) and (e)). The spiral is then unpinned (Figure \ref{fig:unpin}(f)), but will continue to generate new spiral pairs at the obstacle whenever the spiral arm meets it. If the teleporting stimulus extends along the entire contour, the original spiral will be both unpinned and annihilated as it is teleported to the boundary, terminating the reentry.  

To demonstrate the generality of designing stimuli to defibrillate by teleportation, we have successfully applied it using more physiologically accurate cardiac cell models including the 8-variable Beeler-Reuter model \cite{beeler1977reconstruction}, the 19-variable TNNP model \cite{ten2004model}, and the 41-variable OVVR model \cite{o2011simulation}, considered by the FDA to be the most realistic human ventricular model to date. Although the phase space of these models is complicated by their large number of dimensions, suitable indicator functions can still be constructed to properly characterize the level set contour topology \cite{topological}. We find both the $f$ (calcium inactivation) and $h$ (sodium inactivation) gate variables functional substitutions for the generic gate variable in Equation \ref{fg}, while suitable thresholds are identified by inspecting the phase-space portrait as in Figure \ref{fig:spiral}(b). Real-time interactive WebGL programs \cite{kaboudian2019real} for generating defibrillating stimuli in these models are provided at the author's website \cite{abubu}.

\begin{figure}[!ht]
    \centering
    \includegraphics[scale=.22]{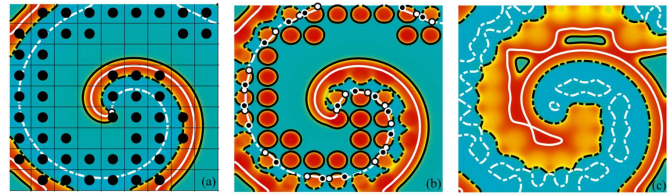}
    \caption{Defibrillation by teleportation using an equally-spaced electrode array. (a) Stimulus pattern generated by electrodes (black) firing for $f<0$ and $|g|<g_\textrm{th}$ in each grid cell. (b) Immediate result of stimulation. Many transient singularities form via the S1-S2 mechanism. (c) Nearby singularities spontaneously annihilate and the original singularity is teleported to the boundary. A complete video is available in SI Appendix, Movie S7.}
    \label{fig:electrodes}
\end{figure}

In order to emulate an experimentally realistic setup, we implemented an array of electrodes from which stimuli may be applied \cite{rappel1999spatiotemporal}. Figure \ref{fig:electrodes}(a) shows an example designed stimulus pattern for an isolated spiral wave in the FHN model. The domain was divided into a $9 \times 9$ array of electrodes. If $f<0$ and $|g|<g_\textrm{th}$ were satisfied within an electrode, a stimulus was applied. Because of the pulses' discontinuity, many new pairs of singularities are created via the S1-S2 mechanism as shown in Figure \ref{fig:electrodes}(b). However, as they are very close, the pairs spontaneously annihilate, producing a continuous excited back as shown in Figure \ref{fig:electrodes}(c). The original singularity is then effectively teleported to the boundary and defibrillation is successful. If the electrodes are insufficiently dense, however, newly created pairs may persist indefinitely and actually increase the total number of singularities.

\section*{Teleportation as the mechanism for defibrillation}
Defibrillation in the heart is actually possible and is very similar to what takes place in the electrodes example, thanks to intramural excitations produced by \textit{virtual electrodes} (VE), which arise due to interactions between tissue heterogeneities and an electric field \cite{efimov1998virtual,fenton2009termination}. The shock-induced teleportation of spiral wave singularities serves as a mechanistic explanation for defibrillation in real tissue at both high \cite{ideker,chattipakorn2004effects} and low energies \cite{li2011low,fenton2009termination}. During high-energy shocks, enough virtual electrodes (VE) are activated to excite the entire heart and thus stimulate continuously along the refractory backs connecting reentrant spiral waves. Pairs of spirals are instantly teleported together, while isolated spirals are teleported to domain boundaries. Low-energy shocks in general do not excite everywhere \cite{fenton2009termination}, but as enough of the refractory back is stimulated with every shock, spirals are teleported closer and closer together until they all annihilate.

Under the theory presented here, two effects can contribute to the failure of high- and low-energy shocks to successfully defibrillate. First, spirals can fail to be teleported close enough to annihilate if their refractory backs are insufficiently stimulated by VE excitations. This may occur during high-energy defibrillation if at least one reentrant pair is not fully teleported together, resulting in persistent rotors which then reinitiate fibrillation \cite{chattipakorn2004effects}. Second, virtual electrodes may discontinuously stimulate refractory backs, resulting in the initiation of new spiral waves as in the S1-S2 mechanism \cite{efimov1998virtual,fenton2009termination}. This explains cases where low-energy defibrillation fails \cite{chattipakorn2004effects,efimov2000direct}. While a single low-energy shock may fail to completely defibrillate, multi-shock therapies such as LEAP may succeed by annihilating more pairs by teleportation in each shock than are created via S1-S2 initiation \cite{leap}. 

\section*{Comparison to other methods}
Spiral waves are commonly found in many excitable systems, from chemical systems such as CO oxidation on platinum, premixed gas flames, and the Belousov–Zhabotinsky (BZ) reaction to biological ones such as slime mold (Dictyostelium discoideum) and Xenopus
laevis oocytes. In the human body, spiral waves can appear not only in the heart, but also on the skin, the tongue, the intestine, the retina, and the brain~\cite{manz2019patterns}. 
Continuing interest in these phenomena has led to the proposal and study of methods for control ~\cite{steinbock1993control,perez1992electric,rappel1999spatiotemporal} and termination of spiral waves both theoretically and experimentally for many years.
These termination methods have been based on slowly moving spiral waves to boundaries by perturbing the spiral wave tips with fast pacing and most recently by perturbing the cores of spiral waves directly using optogenetics~\cite{majumder2018optogenetics,hussaini2021drift}.
By using tissue that can be excited or de-excited optically, a controlled stimulus can be delivered by spots of light and directed at spiral wave cores to affect their dynamics. For example, by using a deactivating (blocking) light-spot, Majumder et al. were able to anchor and then drag a spiral wave across tissue by moving the light spot until annihilation was achieved at a boundary ~\cite{majumder2018optogenetics}. Another approach is to use a gradient of sub-threshold illumination which induces a drift of the spiral wave that can bring it to a boundary to be terminated~\cite{hussaini2021drift}.
 However, not only do all termination methods to date require several periods for spiral waves to be moved across the domain to boundaries or each other, but they only work for stable spiral waves with circular cores. If spiral waves have complex meander or are in an unstable breakup regime, these methods will not succeed as they are too slow to account for the fast creation and annihilation of new waves. In contrast, the methodology presented here allows for the creation of a single stimulus that can be delivered optogenetically or by an array of electrodes that instantaneously will move and extinguish all existing spiral waves in the domain independent of number of spirals (anchored or not) and their dynamics (stable, meandering, or unstable). 

\section*{Conclusions and open questions}
By considering the level set contours organizing excitable dynamics, we have demonstrated a novel minimal defibrillation strategy capable of automatically eliminating any phase singularities responsible for spiral wave turbulence. Using a designed stimulus, pairs of connected spiral wave tips are instantaneously moved close together by a mechanism we term teleportation such that the tips attract and mutually annihilate. The topological nature of this method makes it model-independent and applicable to experiment. A particularly promising experimental testbed would be optogenetically modified cardiac monolayers \cite{optogenetics1,optogenetics2}. Previous studies have shown how rotors in this system may be eliminated using heuristically chosen regions of activation \cite{optogenetics1}. Spatially resolved optical control combined with established methods for determining contours in experiment \cite{robust,robustexp} would allow for exact replication of our computational examples and provide a systematic method of eliminating rotors. Although time delays between mapping and stimulus design are unavoidable, wave velocities in cardiac monolayers are on the order of 20cm/s. Modern GPU calculations for analyzing the contours can easily be performed in the 1-2ms window required for optical mapping, and thus the proposed method is completely feasible. Additionally, the topological specifications are quite robust; stimulation need only be applied in the general vicinity of the refractory boundaries. Even if these regions drift slightly during the computation time, the desired effect of the stimulus is still established.

The mechanism of defibrillation by teleportation easily explains the success and failure of existing defibrillation methods. High-energy shocks automatically stimulate the necessary refractory contours of every spiral pair, while low-energy shocks do not and in general may generate additional spiral pairs. While we have only demonstrated this topological mechanism in 2D systems, it is easy to generalize to 3D, where spiral waves become scroll waves with associated 1D singular filaments. These filaments are organized by refractory and excitable surfaces rather than contours. Just like in 2D, complete stimulation of the refractory surfaces joining filaments will eliminate the associated scroll waves, while partial stimulation results in the initiation of new waves. This generalization could be explored experimentally using the excitable Belousov-Zhabotinsky reaction~\cite{BZ} in 3D. With the necessary topological requirements for defibrillation elucidated, future low-energy defibrillation strategies may be developed that directly address the teleportation mechanism responsible.

% Bibliography
\bibliography{pnas-sample}

\begin{thebibliography}{10}

\bibitem{winfree1972spiral}
AT Winfree, Spiral waves of chemical activity.
\newblock {\em\protect\JournalTitle{Science}} \textbf{175}, 634--636 (1972).

\bibitem{tyson1989spiral}
JJ Tyson, KA Alexander, V Manoranjan, J Murray, Spiral waves of cyclic amp in a
  model of slime mold aggregation.
\newblock {\em\protect\JournalTitle{Physica D: Nonlinear Phenomena}}
  \textbf{34}, 193--207 (1989).

\bibitem{tsuji2019spirals}
K Tsuji, SC M{\"u}ller, {\em Spirals and Vortices}.
\newblock (Springer), (2019).

\bibitem{ricknat}
J Jalife, RA Gray, AM Pertsov, Spatial and temporal organization during cardiac
  fibrillation.
\newblock {\em\protect\JournalTitle{Nature (London)}} \textbf{392}, 75--78
  (1998).

\bibitem{cherry2008visualization}
EM Cherry, FH Fenton, Visualization of spiral and scroll waves in simulated and
  experimental cardiac tissue.
\newblock {\em\protect\JournalTitle{New Journal of Physics}} \textbf{10},
  125016 (2008).

\bibitem{davidenko1992stationary}
JM Davidenko, AV Pertsov, R Salomonsz, W Baxter, J Jalife, Stationary and
  drifting spiral waves of excitation in isolated cardiac muscle.
\newblock {\em\protect\JournalTitle{Nature}} \textbf{355}, 349--351 (1992).

\bibitem{eisenberg1998defibrillation}
MS Eisenberg, Defibrillation: the spark of life.
\newblock {\em\protect\JournalTitle{Scientific American}} \textbf{278}, 86--91
  (1998).

\bibitem{fenton2009termination}
FH Fenton, et~al., Termination of atrial fibrillation using pulsed low-energy
  far-field stimulation.
\newblock {\em\protect\JournalTitle{Circulation}} \textbf{120}, 467 (2009).

\bibitem{efimov1998virtual}
IR Efimov, Y Cheng, DR Van~Wagoner, T Mazgalev, PJ Tchou, Virtual
  electrode--induced phase singularity: A basic mechanism of defibrillation
  failure.
\newblock {\em\protect\JournalTitle{Circulation Research}} \textbf{82},
  918--925 (1998).

\bibitem{li2011low}
W Li, et~al., Low-energy multistage atrial defibrillation therapy terminates
  atrial fibrillation with less energy than a single shock.
\newblock {\em\protect\JournalTitle{Circulation: Arrhythmia and
  Electrophysiology}} \textbf{4}, 917--925 (2011).

\bibitem{leap}
S Luther, et~al., Low-energy control of electrical turbulence in the heart.
\newblock {\em\protect\JournalTitle{Nature (London)}} \textbf{475}, 235--239
  (2011).

\bibitem{rickgray}
RA Gray, N Chattipakorn, HL Swinney, Termination of spiral waves during cardiac
  fibrillation via shock-induced phase resetting.
\newblock {\em\protect\JournalTitle{Proceedings of the National Academy of
  Sciences - PNAS}} \textbf{102}, 4672--4677 (2005).

\bibitem{danwilson}
D Wilson, J Moehlis, Toward a more efficient implementation of antifibrillation
  pacing.
\newblock {\em\protect\JournalTitle{PloS one}} \textbf{11}, e0158239 (2016).

\bibitem{topological}
CD Marcotte, RO Grigoriev, Dynamical mechanism of atrial fibrillation: A
  topological approach.
\newblock {\em\protect\JournalTitle{Chaos (Woodbury, N.Y.)}} \textbf{27},
  093936 (2017).

\bibitem{keener}
JP Keener, The topology of defibrillation.
\newblock {\em\protect\JournalTitle{Journal of theoretical biology}}
  \textbf{230}, 459--473 (2004).

\bibitem{fitzhugh1961impulses}
R FitzHugh, Impulses and physiological states in theoretical models of nerve
  membrane.
\newblock {\em\protect\JournalTitle{Biophysical journal}} \textbf{1}, 445
  (1961).

\bibitem{karma1993spiral}
A Karma, Spiral breakup in model equations of action potential propagation in
  cardiac tissue.
\newblock {\em\protect\JournalTitle{Physical review letters}} \textbf{71}, 1103
  (1993).

\bibitem{experimentalist}
AN Iyer, RA Gray, An experimentalist’s approach to accurate localization of
  phase singularities during reentry.
\newblock {\em\protect\JournalTitle{Annals of biomedical engineering}}
  \textbf{29}, 47--59 (2001).

\bibitem{curvature}
J Beaumont, N Davidenko, JM Davidenko, J Jalife, Spiral waves in
  two-dimensional models of ventricular muscle: Formation of a stationary core.
\newblock {\em\protect\JournalTitle{Biophysical journal}} \textbf{75}, 1--14
  (1998).

\bibitem{fenton1998vortex}
F Fenton, A Karma, Vortex dynamics in three-dimensional continuous myocardium
  with fiber rotation: Filament instability and fibrillation.
\newblock {\em\protect\JournalTitle{Chaos: An Interdisciplinary Journal of
  Nonlinear Science}} \textbf{8}, 20--47 (1998).

\bibitem{robust}
DR Gurevich, RO Grigoriev, Robust approach for rotor mapping in cardiac tissue.
\newblock {\em\protect\JournalTitle{Chaos (Woodbury, N.Y.)}} \textbf{29},
  053101 (2019).

\bibitem{robustexp}
DR Gurevich, C Herndon, I Uzelac, FH Fenton, RO Grigoriev, Level-set method for
  robust analysis of optical mapping recordings of fibrillation.
\newblock {\em\protect\JournalTitle{Computing in Cardiology}} \textbf{44},
  197--427 (2017).

\bibitem{mermin}
ND Mermin, The topological theory of defects in ordered media.
\newblock {\em\protect\JournalTitle{Reviews of modern physics}} \textbf{51},
  591--648 (1979).

\bibitem{winfree}
AT Winfree, {\em When time breaks down : the three-dimensional dynamics of
  electrochemical waves and cardiac arrhythmias / Arthur T. Winfree.}
\newblock (Princeton University Press, Princeton, N.J.), (1987).

\bibitem{winfreetop}
A Winfree, S Strogatz, Singular filaments organize chemical waves in three
  dimensions: I. geometrically simple waves.
\newblock {\em\protect\JournalTitle{Physica. D}} \textbf{8}, 35--49 (1983).

\bibitem{davidsen2004topological}
J Davidsen, L Glass, R Kapral, Topological constraints on spiral wave dynamics
  in spherical geometries with inhomogeneous excitability.
\newblock {\em\protect\JournalTitle{Physical Review E}} \textbf{70}, 056203
  (2004).

\bibitem{pertsov2000topological}
AM Pertsov, M Wellner, M Vinson, J Jalife, Topological constraint on scroll
  wave pinning.
\newblock {\em\protect\JournalTitle{Physical review letters}} \textbf{84}, 2738
  (2000).

\bibitem{noflux}
F Siso-Nadal, NF Otani, RF Gilmour, Jr, JJ Fox, Boundary-induced reentry in
  homogeneous excitable tissue.
\newblock {\em\protect\JournalTitle{Physical review. E, Statistical, nonlinear,
  and soft matter physics}} \textbf{78}, 031925--031925 (2008).

\bibitem{kuramoto}
Y Kuramoto, {\em Chemical oscillations, waves, and turbulence}, Springer series
  in synergetics ; v. 19.
\newblock (none), (1984).

\bibitem{s1s2}
AM Pertsov, JM Davidenko, R Salomonsz, WT Baxter, J Jalife, Spiral waves of
  excitation underlie reentrant activity in isolated cardiac muscle.
\newblock {\em\protect\JournalTitle{Circulation research}} \textbf{72},
  631--650 (1993).

\bibitem{s1s22}
DW Frazier, et~al., Stimulus-induced critical point: mechanism for electrical
  initiation of reentry in normal canine myocardium.
\newblock {\em\protect\JournalTitle{The Journal of clinical investigation}}
  \textbf{83}, 1039--1052 (1989).

\bibitem{mueller2015scotty}
S Mueller, et~al., Scotty: Relocating physical objects across distances using
  destructive scanning, encryption, and 3d printing in {\em Proceedings of the
  Ninth International Conference on Tangible, Embedded, and Embodied
  Interaction}.
\newblock pp. 233--240 (2015).

\bibitem{ideker}
DJ Dosdall, VG Fast, RE Ideker, Mechanisms of defibrillation.
\newblock {\em\protect\JournalTitle{Annual review of biomedical engineering}}
  \textbf{12}, 233--258 (2010).

\bibitem{pumirkrinsky}
A Pumir, VI Krinsky, Two biophysical mechanisms of defibrillation of cardiac
  tissue.
\newblock {\em\protect\JournalTitle{Journal of theoretical biology}}
  \textbf{185}, 189--199 (1997).

\bibitem{maxwell}
A Pumir, VI Krinsky, How does an electric field defibrillate cardiac muscle?
\newblock {\em\protect\JournalTitle{Physica. D}} \textbf{91}, 205--219 (1996).

\bibitem{beeler1977reconstruction}
GW Beeler, H Reuter, Reconstruction of the action potential of ventricular
  myocardial fibres.
\newblock {\em\protect\JournalTitle{The Journal of physiology}} \textbf{268},
  177--210 (1977).

\bibitem{ten2004model}
KH ten Tusscher, D Noble, PJ Noble, AV Panfilov, A model for human ventricular
  tissue.
\newblock {\em\protect\JournalTitle{American Journal of Physiology-Heart and
  Circulatory Physiology}} \textbf{286}, H1573--H1589 (2004).

\bibitem{o2011simulation}
T O'Hara, L Vir{\'a}g, A Varr{\'o}, Y Rudy, Simulation of the undiseased human
  cardiac ventricular action potential: model formulation and experimental
  validation.
\newblock {\em\protect\JournalTitle{PLoS Comput Biol}} \textbf{7}, e1002061
  (2011).

\bibitem{kaboudian2019real}
A Kaboudian, EM Cherry, FH Fenton, Real-time interactive simulations of
  large-scale systems on personal computers and cell phones: Toward
  patient-specific heart modeling and other applications.
\newblock {\em\protect\JournalTitle{Science advances}} \textbf{5}, eaav6019
  (2019).

\bibitem{abubu}
A Kaboudian, Defibrillation by teleportation
  (\url{https://kaboudian.github.io/DefibrillationByTeleportation/}) (2021).

\bibitem{rappel1999spatiotemporal}
WJ Rappel, F Fenton, A Karma, Spatiotemporal control of wave instabilities in
  cardiac tissue.
\newblock {\em\protect\JournalTitle{Physical review letters}} \textbf{83}, 456
  (1999).

\bibitem{chattipakorn2004effects}
N Chattipakorn, I Banville, RA Gray, RE Ideker, Effects of shock strengths on
  ventricular defibrillation failure.
\newblock {\em\protect\JournalTitle{Cardiovascular research}} \textbf{61},
  39--44 (2004).

\bibitem{efimov2000direct}
IR Efimov, Y Cheng, Y Yamanouchi, PJ Tchou, Direct evidence of the role of
  virtual electrode-induced phase singularity in success and failure of
  defibrillation.
\newblock {\em\protect\JournalTitle{Journal of cardiovascular
  electrophysiology}} \textbf{11}, 861--868 (2000).

\bibitem{manz2019patterns}
N Manz, FH Fenton, Patterns and humans.
\newblock {\em\protect\JournalTitle{Spirals and Vortices: In Culture, Nature,
  and Science}} pp. 217--224 (2019).

\bibitem{steinbock1993control}
O Steinbock, V Zykov, SC M{\"u}ller, Control of spiral-wave dynamics in active
  media by periodic modulation of excitability.
\newblock {\em\protect\JournalTitle{Nature}} \textbf{366}, 322--324 (1993).

\bibitem{perez1992electric}
V P{\'e}rez-Mu{\~n}uzuri, R Aliev, B Vasiev, V Krinsky, Electric current
  control of spiral wave dynamics.
\newblock {\em\protect\JournalTitle{Physica D: Nonlinear Phenomena}}
  \textbf{56}, 229--234 (1992).

\bibitem{majumder2018optogenetics}
R Majumder, et~al., Optogenetics enables real-time spatiotemporal control over
  spiral wave dynamics in an excitable cardiac system.
\newblock {\em\protect\JournalTitle{Elife}} \textbf{7}, e41076 (2018).

\bibitem{hussaini2021drift}
S Hussaini, et~al., Drift and termination of spiral waves in optogenetically
  modified cardiac tissue at sub-threshold illumination.
\newblock {\em\protect\JournalTitle{Elife}} \textbf{10}, e59954 (2021).

\bibitem{optogenetics1}
C Crocini, et~al., Optogenetics design of mechanistically-based stimulation
  patterns for cardiac defibrillation.
\newblock {\em\protect\JournalTitle{Scientific reports}} \textbf{6},
  35628--35628 (2016).

\bibitem{optogenetics2}
RAB Burton, et~al., Optical control of excitation waves in cardiac tissue.
\newblock {\em\protect\JournalTitle{Nature photonics}} \textbf{9}, 813--816
  (2015).

\bibitem{BZ}
T Bánsági, O Steinbock, Three-dimensional spiral waves in an excitable
  reaction system: Initiation and dynamics of scroll rings and scroll ring
  pairs.
\newblock {\em\protect\JournalTitle{Chaos: An Interdisciplinary Journal of
  Nonlinear Science}} \textbf{18}, 026102 (2008).

\end{thebibliography}

\end{document}